\documentclass[aps,prd,twocolumn,preprintnumbers,nofootinbib,amsmath,amssymb]{revtex4-2}
\usepackage{graphicx}
\usepackage{dcolumn}
\usepackage{bm}
\usepackage{amsmath}
\usepackage{braket}
\usepackage[hidelinks]{hyperref}
\usepackage{epstopdf}
\usepackage{placeins}
\usepackage{multirow}
\usepackage{makecell}
\usepackage{cleveref}
\usepackage{xcolor}
\usepackage{dblfloatfix}
\usepackage{slashed}

\newcommand{\beq}{\begin{eqnarray}}
\newcommand{\eeq}{\end{eqnarray}}
\newcommand{\beqnn}{\begin{eqnarray*}}
\newcommand{\eeqnn}{\end{eqnarray*}}

\usepackage[font=small, labelfont=bf, textfont={small,it}]{caption}
\def\spose#1{\hbox to 0pt{#1\hss}}
\def\ltapprox{\mathrel{\spose{\lower 3pt\hbox{$\mathchar"218$}}
	\raise 2.0pt\hbox{$\mathchar"13C$}}}

\begin{document}

\title{Topological properties around the Roberge-Weiss transition in $N_f = 2+1+1$ QCD}

\author{Massimo D'Elia}
\email{massimo.delia@unipi.it}
\affiliation{Dipartimento di Fisica dell'Universit\`a di Pisa \& \\ INFN Sezione di Pisa, Largo Pontecorvo 3, I-56127 Pisa, Italy}

\author{Fabio Siliberto}
\email{f.siliberto@studenti.unipi.it}
\affiliation{Dipartimento di Fisica dell'Universit\`a di Pisa \& \\ INFN Sezione di Pisa, Largo Pontecorvo 3, I-56127 Pisa, Italy}

\author{Kevin Zambello}
\email{kevin.zambello@pi.infn.it}
\affiliation{Dipartimento di Fisica dell'Universit\`a di Pisa \& \\ INFN Sezione di Pisa, Largo Pontecorvo 3, I-56127 Pisa, Italy}

\date{\today}

\begin{abstract}
  We investigate the topological properties of QCD across the finite temperature Roberge-Weiss transition, which is found
  for particular values of the imaginary baryon chemical potential.
  Our study is conducted for
  $N_f = 2+1+1$ QCD with physical quark masses, discretized via stout improved staggered fermions and considering
  mostly two different values of the compactifed dimension, $N_t = 8$ and $N_t = 10$. 
  Results for $T_{RW}$ and for the associated universality class are consistent with those found in the
  $N_f = 2+1$ case with a slightly different discretization.
  The topological susceptibility appears to be practically constant for $T \lesssim T_{RW}$,
  then rapidly decaying for higher temperatures. The analysis of the fourth order cumulant of the topological
  charge distribution, $b_2$, reveals that it is compatible with the prediction of the Dilute Instanton Gas Approximation
  right after $T_{RW}$, showing thus a sharp transition, which is more similar to what observed in pure gauge theories
  rather than to full QCD along the standard thermal line, where instead a slower transition was observed in previous
  studies.
\end{abstract}

\maketitle

\section{Introduction}

\label{sec_intro}

Many of the non-perturbative features which characterize QCD and QCD-like theories are related
to the existence of non-trivial topological properties.
Most of these properties are encoded in the 
dependence of the free (vacuum) energy density on the topological parameter $\theta$,
i.e., the coupling to the winding number $Q$
that can be added to the QCD action, $S_{QCD} \to S_{QCD} + i \theta Q$, and
explicitly breaks CP symmetry.
A first-principle investigation of these non-perturbative
properties can be performed through numerical lattice simulations;
however, various analytic approaches permit to obtain predictions
based on certain assumptions or approximations, which turn out to
be reliable within appropriate regimes,
see Refs.~\cite{Vicari:2008jw,Bonanno:2025wcv,bonanno2026topologicalsusceptibilityqcdfinite} for recent reviews
on the subject.

For example, the Dilute Instanton Gas Approximation (DIGA) assumes
that the dominant contributions to the path-integral come from
semiclassical  configurations built around independent (anti)instanton solutions
to the equations of motion, for which $Q$ is integer-valued~\cite{Gross:1980br,Pisarski:1980md}.
Such solutions are localized in space-time and characterized by a typical
size, the instanton radius $\rho$; the model is then based on a 1-loop
computation of the effective single instanton action, where the relevant
coupling is $g(\rho)$, hence reliable only for $\rho \ll \Lambda_{QCD}^{-1}.$
However, the model predicts a size distribution which diverges for large $\rho$,
thus making both the 1-loop computation and the diluteness assumption inconsistent,
unless some independent condition
puts an infrared (IR) cutoff: this is the case for a thermal system,
so that DIGA is expected to provide reliable predictions
at least for $T \gg \Lambda_{QCD}$.

At low-$T$, one can rely on different approximation schemes.
For example, in the presence of light dynamical quarks,
Chiral Perturbation Theory (ChiPT) provides a reliable framework, after moving
the $\theta$-parameter to the light quark matrix by means of the chiral anomaly,
leading the precise quantitative estimates. 
Alternatively, for pure $SU(N)$ gauge theories or when $N \gg N_f$, the large-$N$
expansion provides semi-quantitative reliable predictions, based also on the
assumption of a non-vanishing $\theta$-dependence in the $N \to \infty$ limit,
as required by the Witten-Veneziano mechanism to explain the large value of
the $\eta'$ meson mass.

Lattice simulations have provided a direct test of such predictions.
A remarkable feature, emerging from lattice results,
is that the DIGA approximation seems to provide a reliable description
not only for $T \gg \Lambda_{QCD}$, but already for $T \gtrsim T_c$, where
by $T_c$ we mean either the deconfinement transition temperature (for pure gauge theories)
or the chiral crossover temperature (for full QCD); below $T_c$, instead, $\theta$-dependence
is well described by large-$N$ or ChiPT predictions.
This picture has been clearly established
for pure gauge theories~\cite{Teper:1985gi,Teper:1985ek,Alles:1997qe,Alles:1996nm,DelDebbio:2004vxo,Berkowitz:2015aua, Borsanyi:2015cka,Frison:2016vuc,Kitano:2015fla,Borsanyi:2021gqg,Borsanyi:2022fub,Bonanno:2023hhp},
where additional investigations in trace-deformed theories
have also shown a strict relation between the change of $\theta$-dependence and the
spontaneous breaking of center symmetry~\cite{Bonati:2018rfg,Bonati:2019kmf}.
In full QCD, where center symmetry is explicitly broken by the presence of dynamical
fermions, the picture is defined with a lower level of accuracy,
mostly because of the much larger computational requirements which lead to increased
statistical uncertainties, 
and also because of a less sharp definition of the crossover region itself.
\\

The purpose of the present study is to investigate $\theta$-dependence
in full QCD with physical quark masses,
along a modification of the standard thermal line
which is found in the presence of an imaginary baryon chemical potential and
is known as the Roberge-Weiss (RW) line~\cite{Roberge:1986mm}.
The modification consists in rotating fermionic boundary conditions in the Euclidean time
by an appropriate global phase, for which an exact Ising-like
global symmetry exists that corresponds to a remnant of the pure gauge center symmetry.
This symmetry breaks spontaneously 
at a critical temperature $T_{RW}$, where the so-called RW transition takes place,
which turns out to be in the 3D-Ising universality class,  
with $T_{RW} = 208(5)$~MeV, for QCD at the physical point~\cite{Bonati:2016pwz}.

The properties of the thermal theory around the RW transition recall
many aspects of the standard deconfinement and chiral transitions.
For example, present evidence is consistent with the fact that, in the presence of massless fermions,
chiral symmetry restoration takes place at $T_{RW}$~\cite{Bonati:2018fvg,Cuteri:2022vwk};
moreover, $T_{RW}$ has been shown to coincide with the temperature above which localized modes
in the Dirac spectrum appear~\cite{Cardinali:2021fpu}. More recently, the study of
the RW transition has been proposed as a possible tool to determine the
onset of the conformal window in many-flavor QCD~\cite{DElia:2026oah}.

Because of the presence of a well-defined transition, in particular associated
with the spontaneous breaking of a remnant of the pure gauge center symmetry, it seems
interesting to clarify if the change of $\theta$-dependence, and the onset of a DIGA
regime right above the transition, apply also to this case. To this purpose, we 
investigate the behaviour across $T_{RW}$
of the topological susceptibility and of the quartic cumulant of the topological
charge distribution, $b_2$, which represent good probes to clarify the picture.
We consider $N_f = 2+1+1$ QCD with physical quark masses,
discretized via stout improved staggered fermions, i.e., we take into account
dynamical contributions from up, down, strange and charm quarks.
Dynamical charm contributions are not expected to lead to significant physical effects
for the temperature regime relevant to our study: in fact, our results for $T_{RW}$
and for the universality class of the transition turn out to be consistent with those of Ref.~\cite{Bonati:2016pwz};
nevertheless, given the different discretization adopted, this represents a non-trivial universality test.

The paper is organized as follows. In Section~\ref{setup_sec} we provide details about the discretization
adopted for the partition function and about our determination of the topological charge on gauge configurations.
In Section~\ref{sec_rw} we recall the main features of the RW line and illustrate our results from a finite size scaling
analysis aimed at determining $T_{RW}$ and the universality class of the transition.
In Section~\ref{sec_topo} we illustrate results concerning the topological susceptibility
and the quartic cumulant of the topological charge distribution both
below and above $T_{RW}$. Finally, in Section~\ref{sec_conclusions}, we draw our conclusions.

\section{Numerical set-up}

\label{setup_sec}

We consider the finite temperature partition function of $N_f=2+1+1$ QCD
discretized via stout-improved staggered fermions as follows:
\begin{equation}
  Z=\int{[DU]}\,e^{-S_{YM}} \prod_{f=u,d,s,c}\det{(M_{st,f})}^\frac{1}{4},
\end{equation}
where
\beq
  S_{YM} & = & -\frac{\beta}{3} \, \sum_{\substack{i \\ \mu\ne\nu}}\left(\frac{5}{6}W^{1\times1}_{i,\mu\nu}-\frac{1}{12}W^{1\times2}_{i,\mu\nu}\right) \nonumber \\
{(M_{st,f})}_{ij} & =&  am_f\delta_{ij}+\sum_{\nu=1}^{4}\frac{\eta_{i;\nu}}{2}
\left( e^{a \mu_f \delta_{\nu,4}} U^{(4)}_{i;\nu}\delta_{i\, j-\hat{\nu}} 
 - \right. \nonumber \\
 & &  \left. e^{-a\mu_f \delta_{\nu,4}} U^{(4)\dagger}_{i-\hat{\nu};\nu}\delta_{i\,j+\hat{\nu}}\right)
\eeq
are respectively the tree-level improved Symanzik gauge action~\cite{weisz,curci} and
the staggered Dirac operator ~\cite{kogut-susskind,morningstar}. Here $i, j$ denote lattice sites,
$\mu, \nu$ denote lattice directions, $\beta$ is the inverse gauge coupling, while $W^{1\times1}_{i,\mu\nu}$ and
$W^{1\times2}_{i,\mu\nu}$ are the real parts of the traces of $1\times1$ and $1\times2$ rectangular closed loops.
The coefficients $\eta_{i;\nu}$ are the staggered phases, $U^{(4)}_{i;\nu}$ are four times stout-smeared gauge links,
with isotropic smearing parameter $\rho = 0.125$, $a$ is the lattice spacing, while $m_f$ are the bare quark masses
and $\mu_f$ are the quark chemical potentials associated with each flavor. As better explained
in Section~\ref{sec_rw}, we will consider particular imaginary values of the chemical potentials,
in order to investigate the theory along the so-called Roberge-Weiss line.
The standard rooting trick is used to remove the taste degeneracy of the
staggered Dirac operator. Periodic (antiperiodic) boundary conditions are imposed along the temporal direction for
the bosonic (fermionic) fields.

The quark masses and inverse gauge coupling have been tuned in order to follow a line of constant physics (LCP) derived
by applying a spline interpolation to the scale setting points reported in Ref.~\cite{Borsanyi:2020mff}, which have been
obtained while keeping fixed the hadron masses ratios of $\pi,~K,~\Omega^-$, and the bare quark masses ratios
$m_c/m_s = 11.85$ and $m_d/m_u = 1$.

To investigate the topological properties, we have adopted the following gluonic discretization
of the continuum definition
$Q = \int d^4 x\, q(x)$ where $q(x)$ is the topological charge density $q(x) = {\rm Tr} (G_{\mu\nu} \tilde G_{\mu\nu}) / (16 \pi^2)$:
\begin{equation} Q_L = \sum_n - \frac{1}{2^9 \pi^2} \sum_{\mu,\nu,\sigma,\rho=\pm 1}^{\pm 4} \tilde{\epsilon}_{\mu\nu\sigma\rho} Tr[\Pi_{\mu\nu}(n)\Pi_{\sigma\rho}(n)] \mbox{ , }
\label{eq:topo}
\end{equation}
where $\Pi_{\mu\nu}(n)$ is the plaquette living on the $(\mu,\nu)$ plane at the lattice site $n$ and the Levi-Civita tensor for a negative direction is defined as $\tilde{\epsilon}_{-\mu\nu\sigma\rho} = - \tilde{\epsilon}_{\mu\nu\sigma\rho}$.

It is well known that, like other similar discretizations, this definition is affected by UV fluctuations, which
can be kept under control by applying  a smoothing algorithm configuration by configuration,
then computing the topological charge on the smoothed configurations.
In this work we adopt the cooling method~\cite{BERG1981475,Iwasaki:1983bv,Itoh:1984pr,Teper:1985rb,ILGENFRITZ1986693,Campostrini:1989dh,Alles:2000sc}, which is computationally fast and can be matched to other similar methods, such 
as the gradient flow or smearing, with very good accuracy~\cite{Bonati:2014tqa,Alexandrou:2017hqw,Alexandrou:2015yba}.

After smoothing, the values obtained for $Q_L$ cluster around almost integer values. Therefore, to assign
an integer winding number to each configuration,
we use the definition~\cite{DelDebbio:2002xa,Bonati:2015sqt,Bonanno:2024ggk}:
\begin{equation}
Q = round(\alpha~Q_L^{cool}) \, ,
\label{eq:rounding}
\end{equation}
where the factor $\alpha$ is chosen so that each cluster is assigned to the correct integer value,
$\alpha = min_{x}\langle (round(x~Q_L^{cool}) - x~Q_L^{cool})^2 \rangle$,
which in practice means that $\alpha$ is chosen such that the clusters formed by $\alpha Q_L$
are centered as closely as possible around integer values.

\section{The Roberge-Weiss line}
\label{sec_rw}

The Roberge-Weiss (RW) symmetry is a remnant of center symmetry in full QCD.
For a $SU(N_c)$ lattice gauge theory with periodic boundary conditions in the Euclidean time,
a center transformation consists of multiplying the temporal links on a given time slice by a given
element of the gauge group center, namely, by $\exp( 2\pi i k / N_c)$, $k = 0, \dots, N_c -1$.
This is an exact symmetry of the pure gauge theory, and the Polyakov loop is a possible
order parameter for its spontaneous breaking at the deconfinement transition.
Dynamical fermions break center symmetry explicitly, since the fermion determinants contain a coupling
to the Polyakov loop, like an external field in a spin model,
which for standard thermal boundary conditions favors a real Polyakov loop.

However, when fermionic boundary conditions in the temporal direction are twisted by a common phase $\exp(i \theta_q)$,
like it happens in the presence of an imaginary baryon chemical potential~$\mu_B$~\cite{Alford:1998sd,Lombardo:1999cz,deForcrand:2002hgr,DElia:2002tig,deForcrand:2003vyj,DElia:2004ani,Giudice:2004se,Chen:2004tb,Azcoiti:2005tv},
$\theta_q = {\rm Im} (\mu_B) / (N_c T)$,
the Polyakov loop which enters the determinants is multiplied by $\exp(i \theta_q)$,
and for $\theta_q = (2 k + 1) \pi / N_c$, with $k$ integer, a residual $Z_2$ global symmetry appears,
remnant of the full center group: this corresponds to the so-called RW line. The RW transition
temperature $T_{RW}$ is the point along this line where the residual center symmetry breaks spontaneously.

The RW transition has been investigated in various lattice
studies~\cite{DElia:2007bkz,DElia:2009bzj,Cea:2009ba,deForcrand:2010he,Bonati:2010gi,Cea:2012ev,Wu:2013bfa,Philipsen:2014rpa,Bonati:2014kpa,Wu:2014lsa,Czaban:2015sas,Bonati:2016pwz,Bonati:2018fvg,Philipsen:2019ouy,Cardinali:2021fpu,Cuteri:2022vwk,Brandt:2022jwo,Brandt:2022iwk,Endrodi:2026tqa,Endrodi:2025hlb,DElia:2025ybj,Zambello:2024ucs,DElia:2026oah}.
$T_{RW}$ generally lies above the pseudo-critical crossover temperature, and
it has been found $T_{RW} = 208(5)$~MeV for $N_f = 2+1$ QCD with physical quark masses,
with the transition in the $3D$ Ising universality class.
However, depending on the number of flavors and on the quark masses,
the transition can also be first order or tricritical.
Moreover, the RW transition seems to coincide with the restoration of chiral symmetry
in the massless case~\cite{Bonati:2018fvg,Cuteri:2022vwk}.
\\

\begin{table}[t!]
\centering
\begin{tabular}{|c|c|c|c|}
\hline   Class & $\gamma$ & $\nu$ & $\beta$ \\
\hline        1st Order & 1 & $\frac{1}{3}$ & - \\
\hline        $Z_2$ & 1.2372(5) & 0.6301(4) & 0.3267(10) \\
\hline
\end{tabular}
\caption{Critical exponents for the two possible universality classes of the Roberge–Weiss transition \cite{Blote_1995}.}
\label{tab:exponents}
\end{table}

As mentioned above, the main purpose of the present study is to investigate the topological properties
of QCD along the RW line, with a particular focus on the region around $T_{RW}$. As a preliminary step
in this direction, this section is dedicated to the determination of $T_{RW}$ and of the associated
universality class for $N_f = 2+1+1$ QCD with physical quark masses, adopting the discretization illustrated in previous
section for lattices with a temporal extension $N_t = 8,10$.
Numerical simulations have been performed using the RHMC algorithm~\cite{Kennedy:1998cu,Clark:2006fx}
with the GPU code \texttt{OpenSTaPLE}~\cite{openstaple,Bonati:2017ovw,Bonati:2018wqj}.

Our results, to be illustrated shortly, are consistent, within statistical errors, with those obtained
in Ref.~\cite{Bonati:2016pwz} for $N_f=2+1$ QCD and for the same values of $N_t$. This is
expected, since dynamical charm quark effects are not so relevant at a temperature scale of the order
of 200~MeV; moreover, for these temperatures and for the explored values of $N_t$ the lattice spacing
is of the order of $0.1$~fm, so it is marginally acceptable for a reliable discretization of charm quarks.
Nevertheless, given that we are using a different discretization,
results reported in this section represent a good consistency check for the results
reported in Ref.~\cite{Bonati:2016pwz}.

\begin{table}[t!]
\centering
\begin{tabular}{|c|c|c|c|}
\hline $N_t$ & $N_s$ & $T~[\textrm{MeV}]$ & $a~[\textrm{fm}]$ \\
\hline 8 & 16, 24, 32, 40, 48 & 187–375 & 0.07-0.13 \\
\hline 10 & 20, 30, 36, 40 & 178–287 & 0.07-0.11 \\        
\hline
\end{tabular}
\caption{Summary of explored lattice sizes and temperature ranges. For each lattice size, we have explored
  $O(10)$ different temperatures, with statistics corresponding to $O(10^5)$ unit length RHMC trajectories for each
temperature. For the $16^3 \times 8$ and $20^3 \times 10$
lattices, higher-statistics runs were performed, with $O(10^6)$ trajectories per temperature.
Gauge and fermionic observables have been measured every $1$ and $10$ trajectories, respectively. }
\label{tab:parameters}
\end{table}

Among the various possible and equivalent RW lines, we have chosen the case $\theta_q = \pi$, which 
is like switching the periodic boundary conditions of fermion fields in the Euclidean time direction
from antiperiodic to periodic, as usual for the spatial directions. Indeed, in this case
the RW transition can be interpreted as a finite-size transition, associated with the spontaneous
breaking of charge conjugation~\cite{DeGrand:2006qb,DeGrand:2007tw,Lucini:2007as,Lucini:2009kf}. 
Along this RW line,
suitable order parameters are the imaginary part of the Polyakov loop, as well as the imaginary part of
the quark number density                                                                                                     
\begin{equation}                                                                                                             
n_f = \frac{1}{N_s^3 N_t} \frac{\partial logZ}{\partial \mu_f} \mbox{ . }                                                    
\end{equation}                                                                                                               
Starting from these, one can also define the corresponding susceptibilities,                                                 
\beq                                                                                                                         
\chi_{pol} & = & N_s^3 N_t (\langle |Im(L)|^2\rangle - \langle |Im(L)| \rangle^2) \nonumber \\                               
\chi_f & = & N_s^3 N_t (\langle |Im(n_f)|^2\rangle - \langle |Im(n_f)| \rangle^2) \, .                                 
\eeq                                                               

Fermionic observables are computed using noisy estimators \cite{Dong:1993pk} for the required traces of the Dirac operator. In our calculations we use four $Z_2$ noise sources per gauge configuration.

\FloatBarrier
\begin{figure}[t!]
  \centering
  \includegraphics[width=0.9\linewidth, clip]{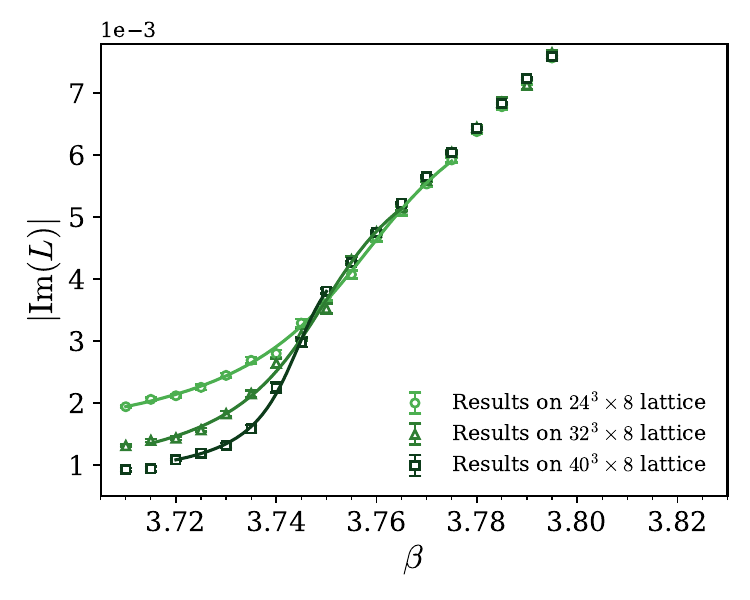}
  \includegraphics[width=0.9\linewidth, clip]{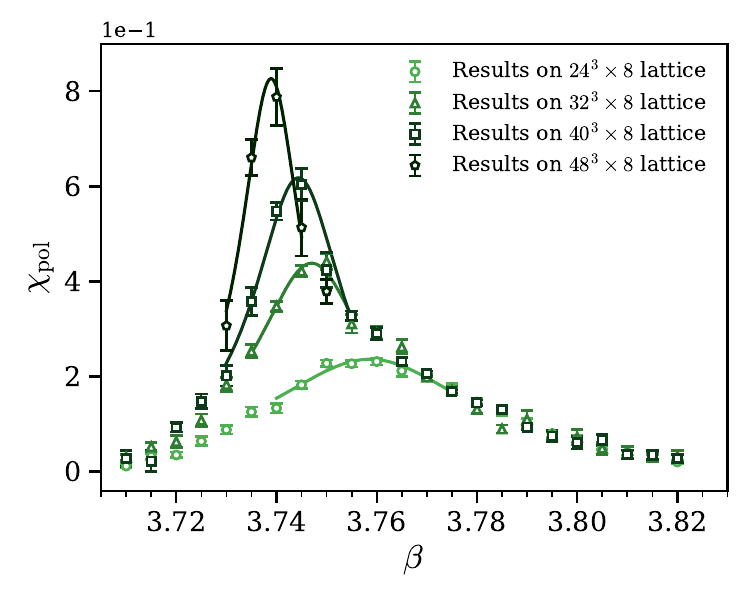}
  \caption{Roberge-Weiss transition: imaginary part of the Polyakov loop (top) and corresponding susceptibility (bottom) for lattices with different volumes and $N_t=8$. The results for the largest lattice ($N_s=48$) are not shown in the top panel because the number of data points is too small for a reliable fit of the Polyakov loop.}
  \label{fig:RW_L_Nt8}
\end{figure}

\begin{figure}[t!]
  \centering
  \includegraphics[width=0.9\linewidth, clip]{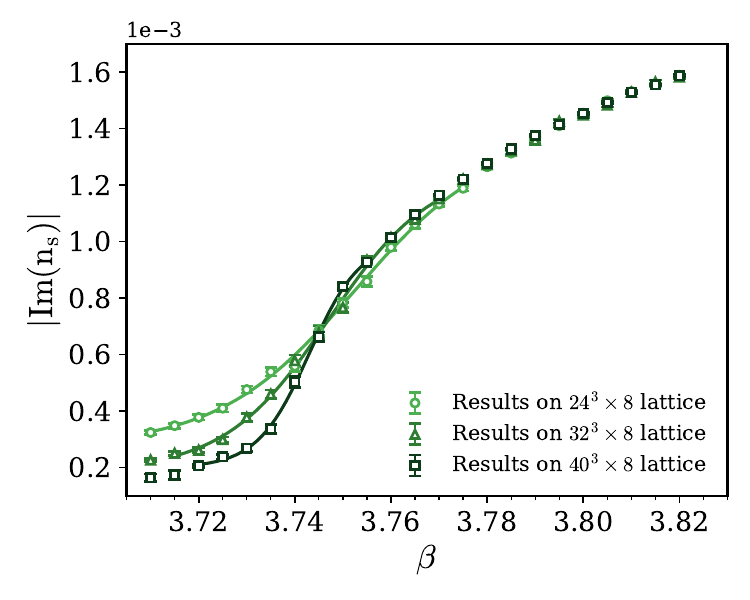}
  \includegraphics[width=0.9\linewidth, clip]{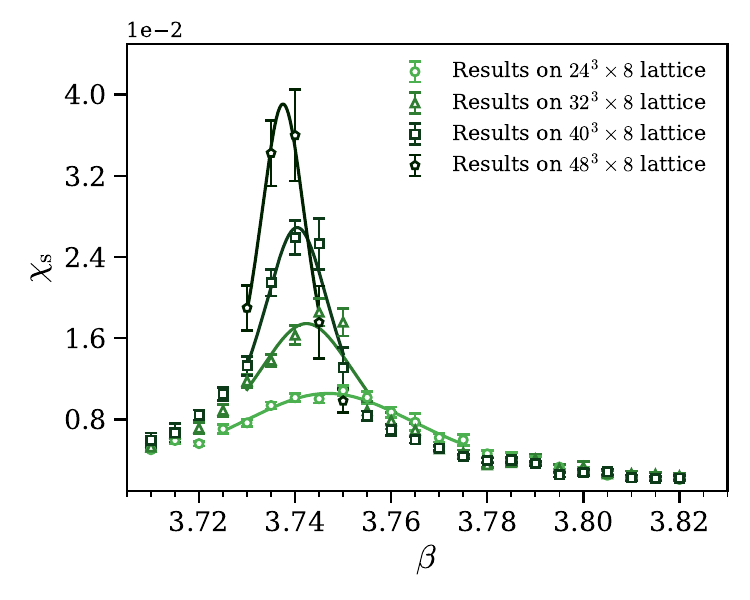}
  \caption{Roberge-Weiss transition: imaginary part of the strange number density (top) and corresponding susceptibility (bottom) for lattices with different volumes and $N_t=8$. The results for the largest lattice ($N_s=48$) are not shown in the top panel because the number of data points is too small for a reliable fit of the strange number density.}
  \label{fig:RW_Ns_Nt8}
\end{figure}

The study of the order parameters and of the peaks of their susceptibilities can be
used locate the transition and to classify the associated critical behavior. Moreover,
from simulations on finite spatial lattice sizes $N_s$, the critical exponents can be obtained
by a finite size scaling (FFS) analysis.
In particular, indicating generically by $m$ and $\chi_m$ 
the order parameter and the corresponding susceptibility, the FSS behavior expected for 
$m$ is
\begin{equation}
m(N_s) = N_s^{\beta / \nu} f[(T-T_{RW}) N_s^{1/\nu}] \mbox{ , }
\label{eq:fss2}
\end{equation}
while that for the susceptibility and for pseudo-critical temperature $T_{RW,p}(N_s)$
\beq
& & \chi_m(N_s) = N_s^{\gamma/\nu} \phi[(T-T_{RW}) N_s^{1/\nu}] \nonumber \\
& & (T_{RW,p}(N_s) - T_{RW}) \sim  N_s^{-1/\nu} \mbox{ . }
\label{eq:fss}
\eeq
Here $f$ and $\phi$ are  universal scaling functions, while $\beta, \gamma$, and $\nu$ are the critical exponents.
In Table~\ref{tab:exponents} we report the known values of such exponents for the 3D-Ising universality class,
which is the one expected to describe a second order transition, given the $Z_2$ symmetry breaking, or the
effective exponents in case of a first order transition.

\subsection{Results on the RW transition for $N_f = 2+1+1$ QCD}

Some relevant parameters used in our simulations, in particular, the different spatial sizes $N_s$,
range of temperatures and corresponding lattice spacings used respectively for $N_t = 8,10$,
are summarized in Tab.~\ref{tab:parameters}. In our simulations, the physical temperature is fixed by the relation
$T = 1 / (N_t a(\beta))$, where $a(\beta)$ is the lattice spacing as a function of the inverse gauge coupling
$\beta$ along the line of constant physics. For this reason, the quantity $(T - T_{RW})$ appearing
in the FSS relations in eqs.~(\ref{eq:fss2}) and (\ref{eq:fss}), is traded for
$\beta - \beta_{RW}$, where $\beta_{RW}$ is the critical value of $\beta$ for the given $N_t$,
since $(T-T_{RW})$ is proportional to $(\beta-\beta_{RW})$ sufficiently close to the critical point.

The top panel of Fig.~\ref{fig:RW_L_Nt8} illustrates the modulus of the imaginary part of the Polyakov loop
as a function of $\beta$ for various volumes and $N_t=8$. As expected, the observable is close to zero at low $T$,
while it acquires a non-zero value at high $T$, thereby signalling the spontanous breaking of the
$Z_2$ RW symmetry.
We estimate the pseudo-critical coupling $\beta_{RW,p}(N_s)$
by fitting the data to atan-like and tanh-like functions.

Additionally, we calculate the pesudo-critical couplings
also from the peaks of the corresponding susceptibility,
as depicted in the bottom panel of Fig.~\ref{fig:RW_L_Nt8}: in this case,
in order to deal with the presence of a non-trivial background and
enhance the visibility of the peaks, so as to better identify the critical behavior,
we have substracted the susceptibility computed on the smallest volume ($N_s=16$).
The position of the peaks is then identified by fitting the subtracted susceptibilities
to a Lorentzian function.

\begin{figure}[t!]
  \centering
    \includegraphics[width=0.9\linewidth, clip]{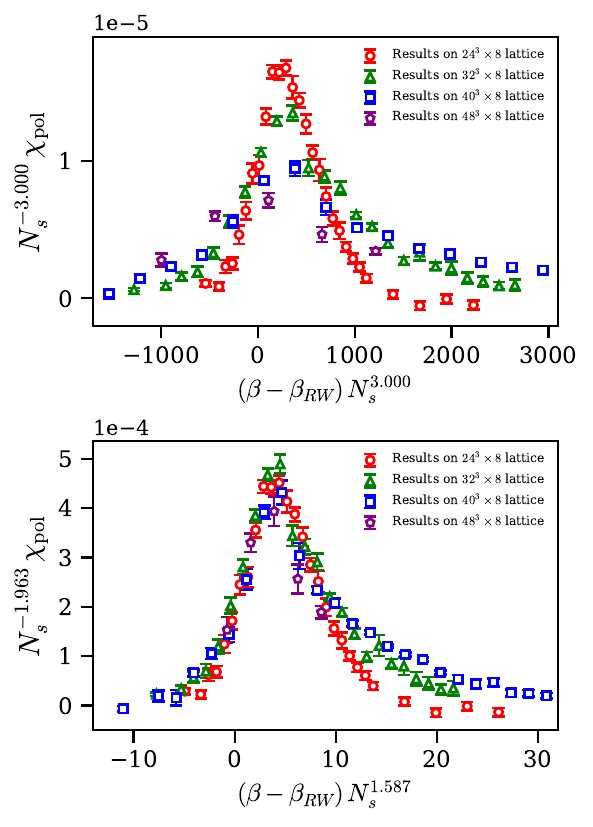}
  \caption{Roberge-Weiss transition: collapse plots of the susceptibility of the Polyakov loop using first-order (top) and 3D Ising (bottom) critical exponents for $N_t=8$ lattices.}
  \label{fig:RW_ChiL_FSS_Nt8}
\end{figure}

In both cases, the errors on the pseudocritical couplings have been estimated taking also into account
the variability observed while varying the fitted range or the fitting function.
A similar analysis has been repeated
for the strange quark number density and its susceptibility, as illustrated in Fig. \ref{fig:RW_Ns_Nt8},
even if in this case we did not need to subtract any background from the susceptibilities.

As a consequence of eqs.~(\ref{eq:fss2}) and (\ref{eq:fss}), we expect that if we plot the
rescaled order parameter or susceptibility (respectively by $N_s^{\beta / \nu}$ and
$N_s^{-\gamma / \nu}$) as a function of $(\beta-\beta_{RW})N_s^{1/\nu}$,
the data from different volumes should collapse onto a single universal curve
if the correct critical exponents and $\beta_{RW} (N_t)$ are used: this is the standard way
to check a given ansatz on the critical behavior by a FSS analysis.

\begin{figure}[t!]
  \centering
  \includegraphics[width=0.9\linewidth, clip]{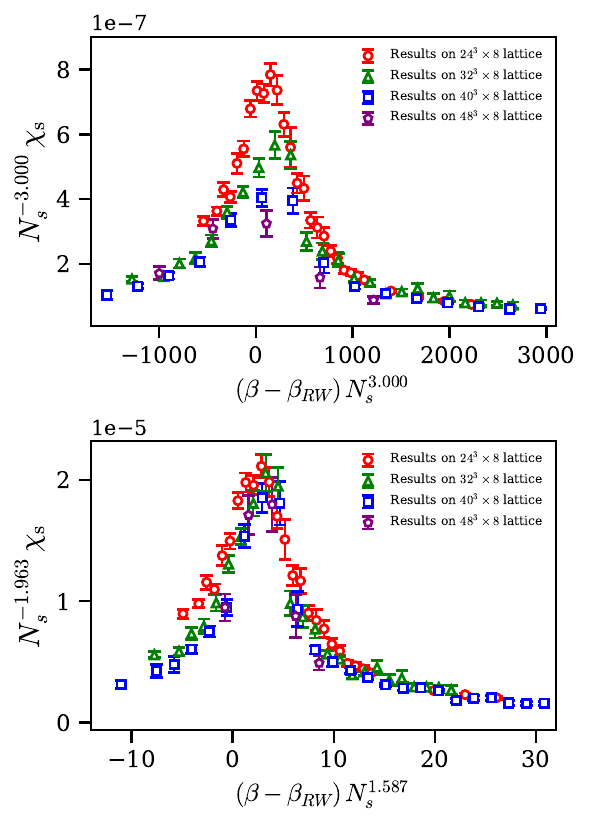}
  \caption{Roberge-Weiss transition: collapse plots for the susceptibility of the strange number density using first-order (top) and 3D Ising (bottom) critical exponents for $N_t=8$ lattices.}
  \label{fig:RW_Chis_FSS_Nt8}
\end{figure}

\begin{figure}[t!]
  \centering
  \includegraphics[width=0.9\linewidth, clip]{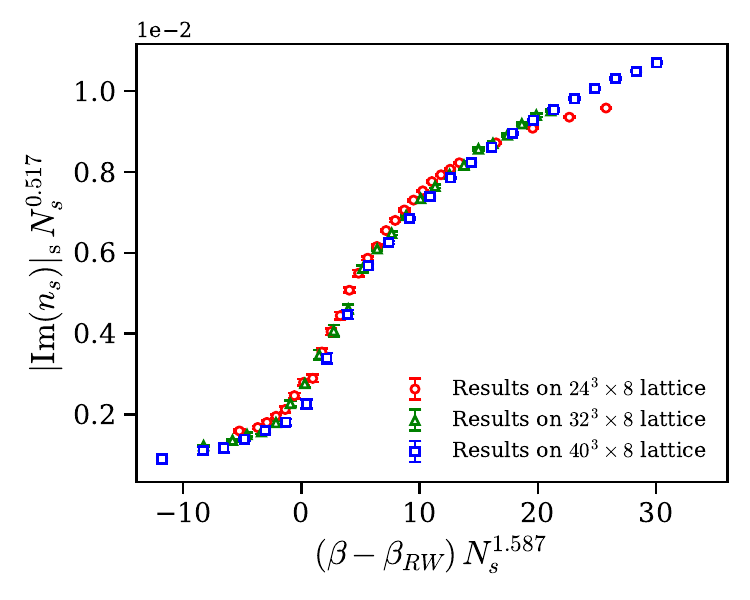}
  
  \caption{Roberge-Weiss transition: collapse plots for the strange number density using the 3D Ising critical exponents for $N_t=8$ lattices.}
  \label{fig:RW_snum_FSS_Nt8}
\end{figure}

Figs.~\ref{fig:RW_ChiL_FSS_Nt8} and \ref{fig:RW_Chis_FSS_Nt8}
illustrate the collapse plots,  for the susceptibility of the Polyakov and the susceptibility of the strange quark number density respectively, using first-order (top) and 3D-Ising (bottom) critical exponents.
Fig.~\ref{fig:RW_snum_FSS_Nt8} also shows the collapse plot for the strange quark number density using 3D-Ising
critical exponents. Such plots provide enough evidence to exclude a first order transition, confirming at the same
time that the RW transition with physical quark masses is 
consistent with the 3D-Ising universality class, as found in
Ref.~\cite{Bonati:2016pwz}.

\begin{figure}[ht!]
  \centering
  \includegraphics[width=0.9\linewidth, clip]{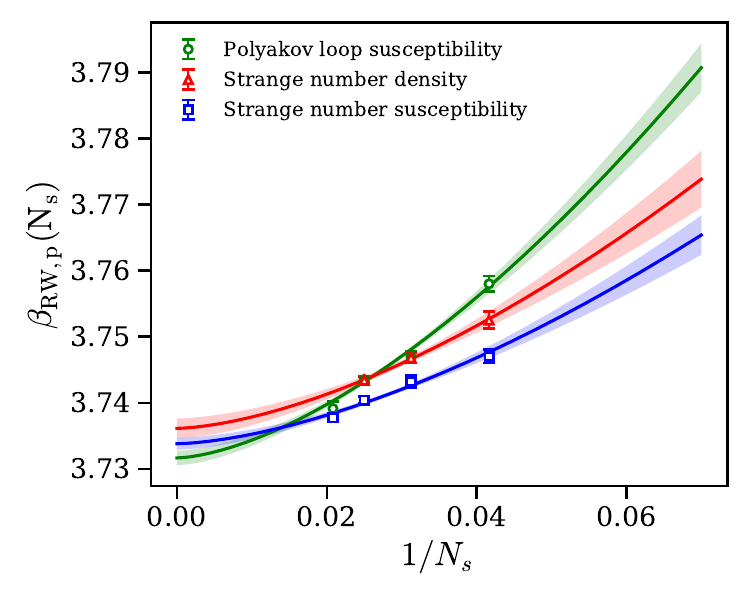}
  \includegraphics[width=0.9\linewidth, clip]{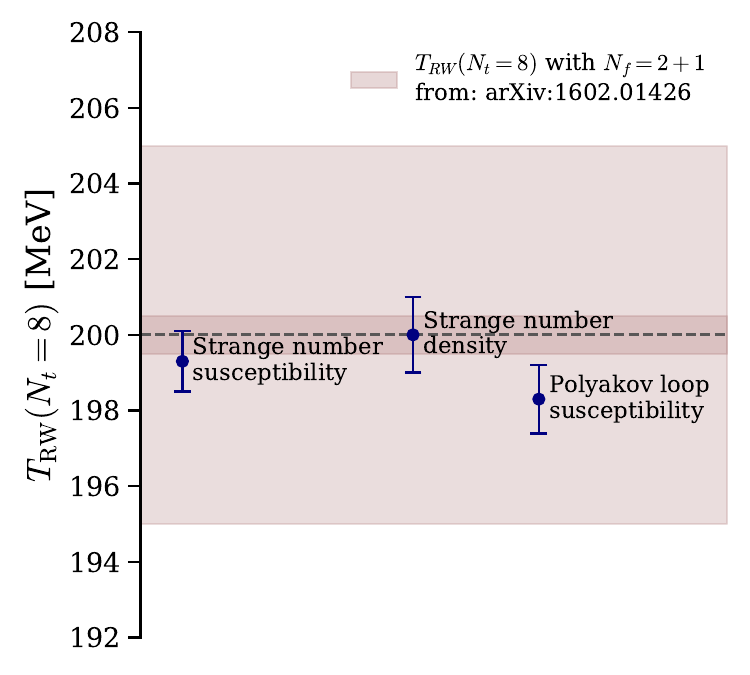}
  \caption{Roberge-Weiss transition: thermodynamic limit extrapolation of $T_{RW}$ for $N_t=8$ lattices (top). Different colors represent extrapolations from different observables. Determination of $T_{RW}$ from different observables in the thermodynamic limit (bottom). The colored bands indicate the results by Ref. ~\cite{Bonati:2016pwz}, shown both with (wider band) and without (narrower band) the systematic uncertainty arising from the determination of the physical scale.}
  \label{fig:RW_Vinf_tT8}
\end{figure}

Having established the order of the transition,
we can use the second of Eqs.~(\ref{eq:fss}) to extrapolate the Roberge-Weiss temperature in the thermodynamic limit.
This is shown in the top panel of Fig.~\ref{fig:RW_Vinf_tT8}, where different colors represent extrapolations derived
from different observables, which agree among each other within $2\sigma$, thus confirming the consistency
of our analysis. The extrapolated values are converted in terms of $T_{RW}$ in the bottom panel
Fig.~\ref{fig:RW_Vinf_tT8}), where systematics related to the determination of the physical scale
have been added; taking into account the systematics from the different observables, we find $T_{RW} = 199.0 \pm 1.4$~MeV. In the same picture we also report the critical temperature obtained for $N_f = 2+1$ QCD and for the same value of $N_t$ in Ref.~\cite{Bonati:2016pwz}, showing very good agreement within errors.

\begin{figure}[ht!]
  \centering
  \includegraphics[width=0.9\linewidth, clip]{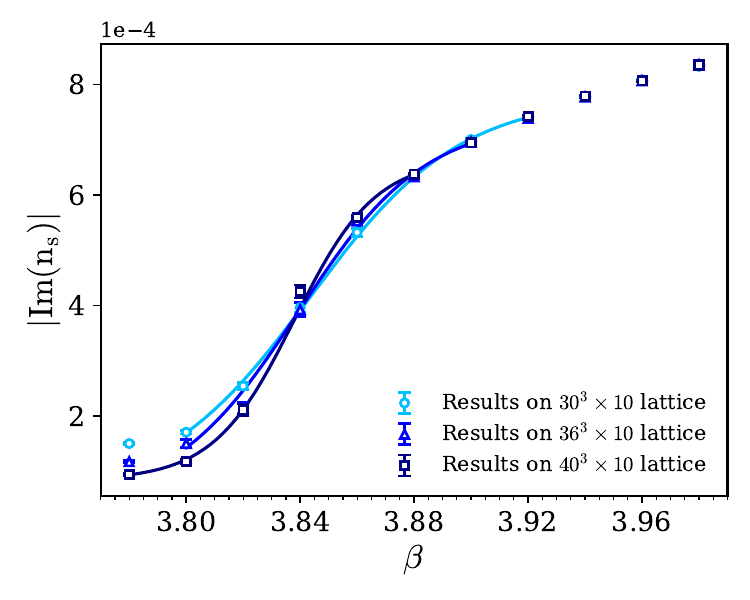}
  \includegraphics[width=0.9\linewidth, clip]{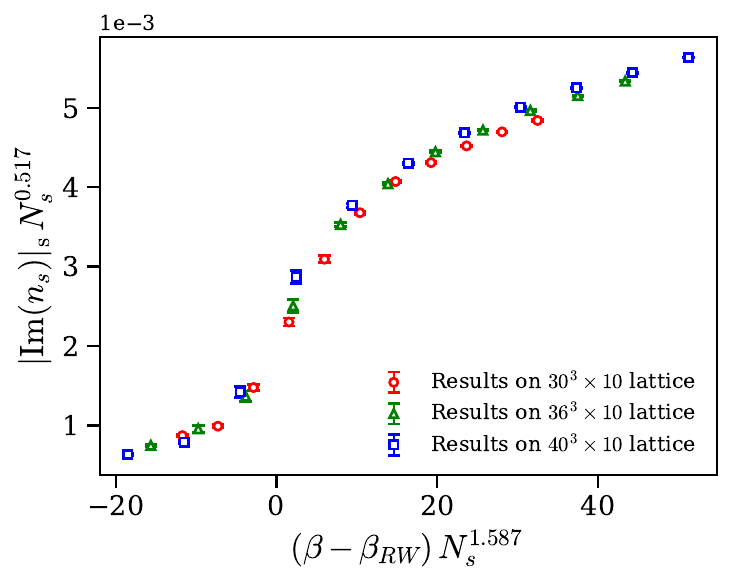}
  \caption{Roberge-Weiss transition: imaginary part of the strange number density (top) for lattices with different volumes and $N_t=10$. Also shown is the collapse plot for the strange number density using the 3D Ising critical exponents (bottom).}
  \label{fig:RW_Ns_Nt10}
\end{figure}

To verify that these findings are stable as we approach the continuum limit, we have repeated a similar analysis
using  $N_t=10$ lattices. Due to limited statistics, in this case our results provide a clear signal only for the
strange quark number density, which is reported in the top panel of Fig.~\ref{fig:RW_Ns_Nt10}.
Using this observable we calculate the Roberge-Weiss temperature, extrapolated to the thermodynamic limit,
yielding $T_{RW} = 204 \pm 7$~MeV. This value is again compatible with what has been reported
in Ref.~\cite{Bonati:2016pwz} for the same value $N_t$. In the bottom panel of Fig.~\ref{fig:RW_Ns_Nt10} we also
report a collapse plot according to Eq.~(\ref{eq:fss}), which shows a good agreement with the 3D-Ising
universality class also in this case.

\section{Topological properties along the RW line}

\label{sec_topo}

Most of the information regarding topological properties is contained in the dependence
of the free-energy density $f(T,\theta)$ on the topological parameter $\theta$.
Since $f(T,\theta)$ is even in $\theta$ and analytic around $\theta = 0$~\cite{Vafa:1984xg},
it is customary to consider the following parameterization
\beq
\label{eq:thetadep_free_energy_general}
f(T,\theta) - f(T,0) = \frac{1}{2} \, \chi(T) \, \theta^2 \hspace{-2pt} \left(\hspace{-2pt}
1+\sum_{n\,=\,1}^{\infty}b_{2n}(T)\theta^{2n} \hspace{-2pt} \right) \hspace{-3pt} .
\eeq
Such parameterization is useful because, given that direct numerical simulation a $\theta \neq 0$
are hindered by the sign problem (the Euclidean action becomes complex), the expansion coefficients
can be related to the topological charge distribution $\mathcal{P}(Q)$ at $\theta = 0$. Indeed, the 
leading order $\mathcal{O}(\theta^2)$ coefficient is the {topological susceptibility}:
\beq
\chi = \frac{\braket{Q^2}}{V}\Bigg\vert_{\theta \, = \, 0},
\eeq
where $V$ is the four-dimensional volume, while higher-order
coefficients, which parameterize the deviations of
$\mathcal{P}(Q)$ from a pure Gaussian, are {related to} higher-order cumulants, denoted as $\braket{Q^k}_c$:
\beq\label{eq:b2n_def}
b_{2n} = \frac{2(-1)^{n}}{(2n+2)!} \frac{\braket{Q^{2n}}_{c}}{\braket{Q^2}}\bigg\vert_{\theta\,=\,0} \, ;
\eeq
for example we have 
\beq
b_2 = -\frac{1}{12} \frac{\braket{Q^4} - 3\braket{Q^2}^2}{\braket{Q^2}} \, .
\eeq
In QCD with dynamical fermions, reliable estimates for $f(T,\theta)$ at $T \simeq 0$
come from ChiPT~\cite{DiVecchia:1980yfw,GrillidiCortona:2015jxo}, in particular,
for the isospin-symmetric case (equal up and down masses), the prediction is~\cite{GrillidiCortona:2015jxo}
$\chi^{1/4} = 77.8(4)$~MeV and $b_2 = -0.022(1)$. Instead, for high temperatures
the DIGA prediction is expected to provide a reliable description~\cite{Gross:1980br,Pisarski:1980md}
\beq
f(T,\theta) - f(T,0) = \chi(T) (1-\cos \theta) \ ,
\eeq
where the dependence on $\theta$ is uniquely fixed by the hypothesis that the relevant field configurations
are a superposition of non-interacting integer charged objects, while the dependence on $T$ can be further
predicted based on a semiclassical approximation~\cite{Gross:1980br,Pisarski:1980md,Boccaletti:2020mxu}: 
\beq
\chi(T) \propto T^{4 - \frac{11}{3} N_c  - \frac{1}{3}N_{f}} \, .
\label{eq:asym_chi_DIGA}
\eeq
At $T \simeq 0$,
various lattice studies have confirmed the ChiPT prediction
for $\chi$ in full QCD~\cite{Bonati:2015vqz,Borsanyi:2016ksw,Alexandrou:2017bzk,Athenodorou:2022aay}
while no reliable numerical estimates still exist for $b_2$.
Along the standard thermal line,
it is generally observed
that $\theta$-dependence remains stable in the low-temperature phase,
while various lattice studies
have shown that $\chi$ rapidly drops above the chiral
crossover~\cite{Alles:2000cg, Bonati:2015vqz,Borsanyi:2016ksw,                                                              
Petreczky:2016vrs, Bonati:2018blm,                                                                                          
Burger:2017xkz,Athenodorou:2022aay,Chen:2022fid, Kotov:2025ilm}.  Also the
quartic coefficient $b_2$ exhibits a fast approach to the DIGA value, $b_2 = -1/12 \simeq -0.0833$, above the
chiral crossover, although a bit slower than in pure
Yang--Mills~\cite{Bonati:2013tt,Bonati:2015vqz,Kotov:2025ilm}.
\\

In this section, we investigate the behaviour of the topological susceptibility and of the $b_2$
cumulant along the RW line, in particular across the RW transition, to see if any significant
differences emerge with respect to the standard thermal theory.

For the topological charge, we adopt the gluonic definition given in Eq.~(\ref{eq:topo}) on configurations smoothed
by a certain number $n_c$ of cooling steps, defining $Q$ by the rounding procedure in Eq.~(\ref{eq:rounding}).
The primary effect of smoothing is to rapidly suppress renormalization effects, yielding almost quantized values of $Q_L$
and, as a consequence of that, a well defined assignment of the topological sector $Q$, via Eq.~(\ref{eq:rounding}).
However, since lattice configurations with non-trivial topology are tipically metastable under action minimization,
a secondary effect is the slow loss of topological signal: this effect can be monitored and
taken into account by a suitable zero-smoothing extrapolation. Since such effects are generally
known to be quadratic, at the leading order, in the smoothing radius $r_s$, and since $r_s \propto \sqrt{n_c}$,
our determinations of $\chi$ and $b_2$ are based on a linear extrapolation in $n_c$, performed in
suitable ranges of $n_c$ where renormalization effects have already been suppressed.

\begin{figure}[t!]
  \centering
  \includegraphics[width=0.9\linewidth, clip]{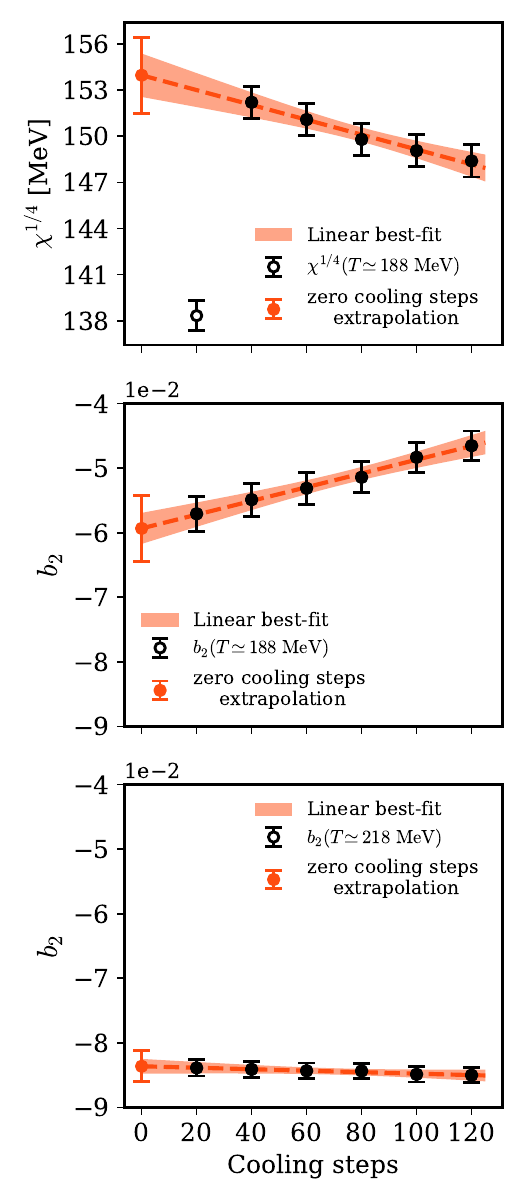}
  \caption{Details on the zero-smoothing extrapolation for $\chi$ and $b_2$, plotted
    as a function of $n_c$ in some example cases, together with a linear best fit and the extrapolated results.
    Top: topological susceptibility at $T \simeq 188$~MeV on the $40^3\times 8$ lattice.
    Middle: $b_2$ cumulant at $T \simeq 188$~MeV on the $16^3 \times 8$ lattice.
        Bottom: $b_2$ cumulant at $T \simeq 218$~MeV on the $16^3 \times 8$ lattice.}
  \label{fig:zerosmoothing_extr}
\end{figure}

In Fig.~\ref{fig:zerosmoothing_extr} we show the details of such procedure in some example cases, both
for $\chi^{1/4}$ and $b_2$.
The slope of $\chi^{1/4}$ as a function of $n_c$ is always negative, independently of $T$, because the long-term
effect of smoothing is always a loss of topological signal, which reduces the variance of $Q$. Instead, for $b_2$
we observe two different behaviors, with a positive or negative slope
depending on whether $T$ is below or above $T_{RW}$; actually, $b_2 (n_c)$ is almost
flat in the latter case. The final errors quoted for the extrapolated data take into account the typical statistical
error of the single data point used in the fit (data entering in each fit are strongly correlated to each other)
and the observed variability of the extrapolation, depending on the choice of the fitted range.

\subsection{Numerical results on the topological susceptibility}

Regarding the topological susceptibility, we are mostly interested in the following two questions:
{\em i)} is $\chi$ approximately constant in the region below $T_{RW}$ and compatible with the value
it takes\footnote{It should be noted that the reference $T = 0$ theory is the same 
for both the standard thermal line and the RW line, since the different boundary conditions
for quark fields in the Euclidean time direction become irrelevant in the $T \to 0$ limit.}  at $T = 0$?
{\em ii)} does $\chi$ drop according to a power law in $T$ above $T_{RW}$, and how does the power law compare
with the DIGA prediction or with what observed along the standard thermal line?

\begin{figure}[t!]
  \includegraphics[width=1.0\linewidth, clip]{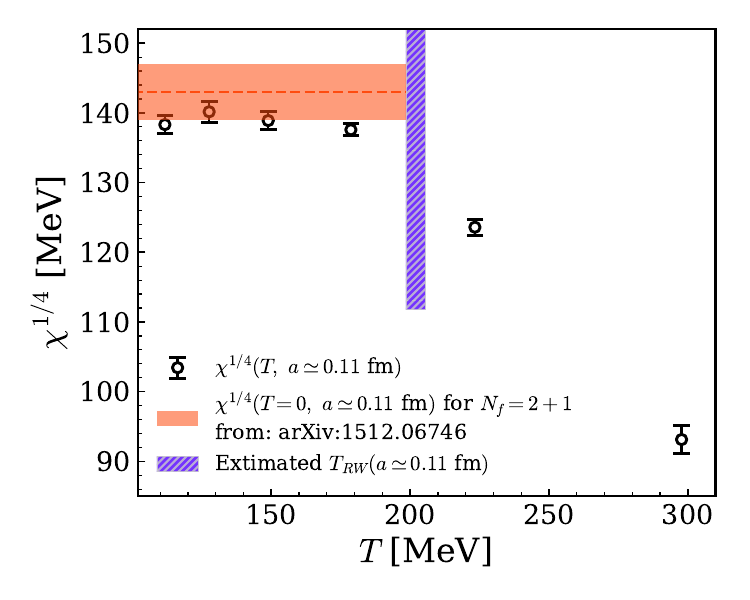}
  \caption{Fourth root of the topological susceptibility as function of the temperature, computed at fixed lattice spacing $a \simeq 0.11~\text{fm}$. The temporal lattice sizes span from $N_t = 6$ to $N_t = 16$. The horizontal band shows the low temperature prediction at a similar lattice spacing extrapolated from Ref.~\cite{Bonati:2015vqz}, while the vertical band
    shows the position of $T_{RW}$ expected from an interpolation of our determinations obtained on
    $N_t = 8$ and $N_t = 10$ lattices.}
  \label{fig:topo_fixed}
\end{figure}

In both cases, in order to reach a reliable answer, one must take into account the strong discretization effects
which affect the determination of $\chi$ in full QCD with physical quark masses~\cite{Bonati:2015vqz,Borsanyi:2016ksw}.
The fermion determinant plays a relevant role in suppressing the weight of configurations with $Q \neq 0$, because
of the associated zero modes, hence the resulting strong suppression of $\chi$ with respect to the quenched case,
with $\theta$-dependence disappearing at all as some of the quark masses vanish. However, for usual discretizations
of the Dirac operator, like the staggered one adopted in the present investigation, zero modes are correctly visible
only as the continuum limit is approached: the consequence is a largely inefficient suppression of topological
fluctuations by the fermion determinant, with resulting large discretization effects which make $\chi$ a rapidly decreasing
function of the lattice spacing $a$.

Therefore, in order to determine whether $\chi(T)$ is approximately constant below $T_{RW}$, then dropping
above it, we must compare determinations at different temperatures but keeping the
lattice spacing $a$ fixed, i.e., we must follow a fixed scale approach in which $T = 1/(N_t a)$
is changed by varying $N_t$ at fixed $a$.
To this purpose, we have performed a set of simulations on $40^4 \times N_t$ lattices, with $N_t = 6,8,10,12,14,16$, and
$a \simeq 0.11$~fm, corresponding to $T \in [110-300]$~MeV, whose results (in terms of $\chi^{1/4}$)
are reported in Fig.~\ref{fig:topo_fixed}.
In the same figure, we also report the estimated location of $T_{RW}$, which is an interpolation of the results
reported above respectively for $N_t = 8$ and $N_t = 10$. As far as we can see, the topological susceptibility is
approximately constant below $T_{RW}$, and drops above it: given the low temperature resolution of this
fixed-scale approach, we cannot resolve the region around $T_{RW}$ more precisely, so it is not excluded that the
drop of $\chi$ starts a bit earlier than $T_{RW}$.

It is interesting to compare the values of $\chi$ below $T_{RW}$ with the $T \simeq 0$ determination.
We cannot compare directly with the ChiPT prediction, because of the strong discretization effects mentioned above.
On the other hand, we have not performed a dedicated zero temperature simulation at the same lattice spacing.
For these reasons, we have considered the determinations of $\chi$ at $T \simeq 0$ reported in Ref.~\cite{Bonati:2015vqz}
for $N_f = 2+1$ QCD with physical quark masses, thereby obtaining an estimate for $a \simeq 0.11$~fm which
is reported for comparison in Fig.~\ref{fig:topo_fixed}: the agreement is reasonable, given also that, although
both based on stout-staggered fermions, the two discretizations adopted in this study and in Ref.~\cite{Bonati:2015vqz}
are slightly different.
\\

\begin{figure}[t!]
\includegraphics[width=1 \linewidth, clip]{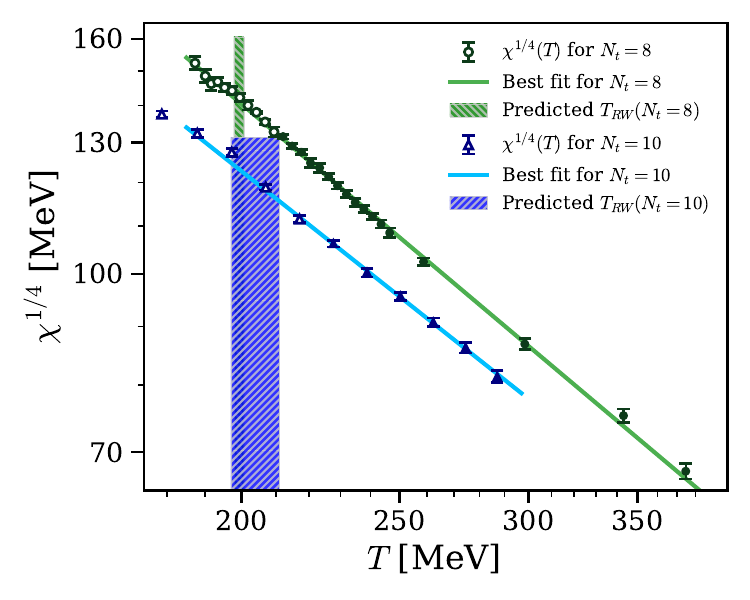}
  \caption{Fourth root of the  topological susceptibility as a function of temperature, on bi-logarithmic scale, computed on the $40^3 \times 8$ and $40^3 \times 10$ lattices. The solid lines show the power-law like best fits obtained using the data above the corresponding phase transitions (filled symbols). The locations of the phase transitions are indicated by the corresponding vertical bands.}
  \label{fig:topo_continuum}
\end{figure}

In order to better study the drop of $\chi$ above $T_{RW}$, we have switched to a
fixed-$N_t$ approach. Fig.~\ref{fig:topo_continuum} illustrates $\chi^{1/4}$ as a function of $T$,
measured on $40^3 \times 8$ and   $40^3 \times 10$ lattices.
Data have been reported in logarithmic scale for both axes, so that
a power law drop $\chi \propto T^c$, as predicted by DIGA, is better visible as a linearly decreasing behaviour.
However, a power law fit $\chi^{1/4} = A T^{c/4}$ returns $c \simeq -4.7$ for $N_t = 8$ an $c \simeq -4.4$ for
$N_t = 10$. For the fixed-$a$ approach, only two data points are available above $T_{RW}$ due to the lower granularity relative to the fixed-$N_t$ approach. Nevertheless, a rough estimate of the power law coefficient can be obtained from these points, yielding $c \sim -3.9$. These estimates are lower than the 
DIGA prediction in Eq.~(\ref{eq:asym_chi_DIGA}) which, considering $N_f = 2,3$ light flavors,
leads to  $c\sim -7.7, -8$: this is expected on the basis of residual lattice spacing
effects~\cite{Bonati:2015vqz,Borsanyi:2016ksw}. However, since our purpose is to just compare
the behavior observed along the RW line with that along the standard thermal line, and we do not aim at a reliable
continuum extrapolation, it is sufficient to observe that the power law coefficient is compatible with
that reported in Ref.~\cite{Bonati:2015vqz}, where similar values of $N_t$ were used.

\subsection{Numerical results on the fourth-order cumulant}

We conclude by presenting the results for the fourth cumulant $b_2$. The interest in this quantity is that
its approach to the value $b_2 = -1/12$, predicted by the $\cos \theta$-like behaviour of the free energy
density, is a distinctive feature of the DIGA model, which is solely related to the absence of interactions
among the topological excitations, hence more essential that the power law behavior of the topological susceptibility,
which instead relies on a semiclassical calculation.

The fourth-order cumulant is notoriously a challenging observable, in particular from the point of view
the achievable statistical uncertainties, which worsen as the volume is increased.
This is a well known problem which affects the measurement of fluctuations of observable, see
for instance the discussion in Refs.~\cite{Bonati:2015sqt,Bonati:2016tvi}. In a few words, the determination
of $b_2$ is based on $\langle Q^4 \rangle_c$, which measures
the deviation from a Gaussian of the topological charge distribution: such deviations are more and more difficult
to detect as the volume increases. 
For this reason, early determinations of $b_2$  were affected by 
large statistical error even in the $SU(3)$ pure gauge
theory~\cite{DelDebbio:2002xa,DElia:2003zne,Giusti:2007tu}. A substantial
improvement, leading eventually to achieve a $O(1\%)$ precision on $b_2$ 
was obtained in Refs.~\cite{Bonati:2015sqt,Bonati:2016tvi}, based on analytic continuation
from simulations at imaginary values of $\theta$~\cite{Panagopoulos:2011rb}.

Unfortunately, due to renormalization issues, the extension to full QCD of analytic continuation from
simulations at imaginary $\theta$ is non-trivial; this is one of the reason why a reliable determination of
$b_2$ at $T = 0$ is still lacking. In order to partially cope with this problem, we decided to adopt a strategy
based on the combination of large statistics 
and intermediate volumes. In particular, we performed a dedicated set of simulations
on lattices with aspect ratio two, $16^3 \times 8$ and $20^3 \times 10$: the possible impact of
finite size effects on such lattices will be discussed shortly.

\begin{figure}[t!]
  \centering
  \includegraphics[width=1.0\linewidth, clip]{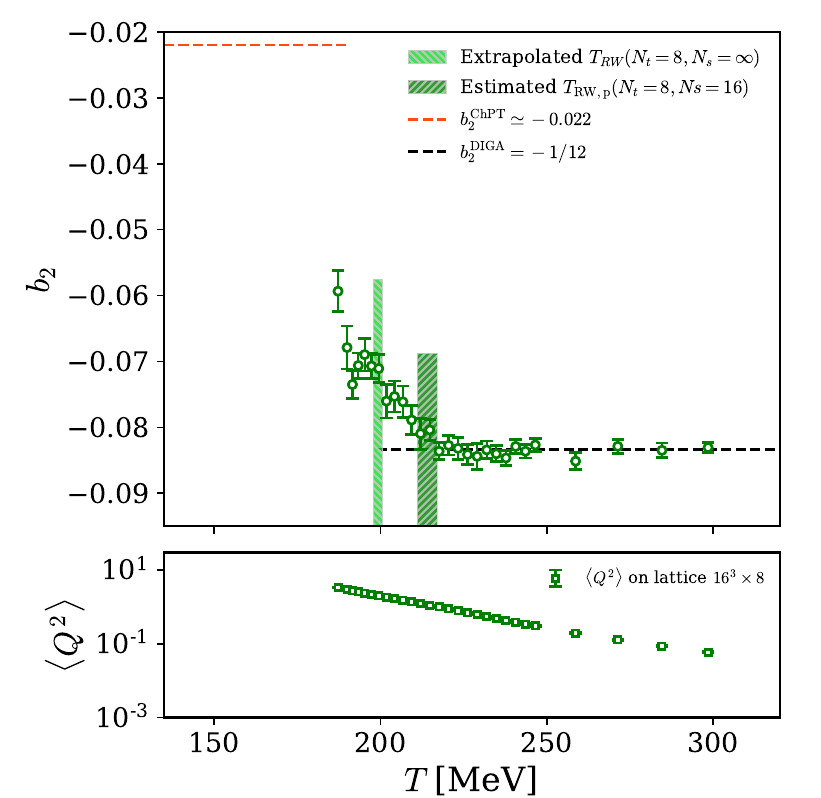}
  \caption{The fourth-order cumulant $b_2$ and $\braket{Q^2}$ as functions of temperature, computed on the $16^3 \times 8$.
        The vertical bands indicates $T_{RW}$ extrapolated to the thermodynamical limit, together
      with an estimate on the same lattice adopted for this measurement,
    while the horizontal dashed lines show the predicted values
    of the cumulant at low and high temperatures, respectively by ChiPT and DIGA. $\braket{Q^2}$ is here shown on a logarithmic scale.}
  \label{fig:b2_nt8}
\end{figure}

\begin{figure}[t!]
  \centering
  \includegraphics[width=1.0\linewidth, clip]{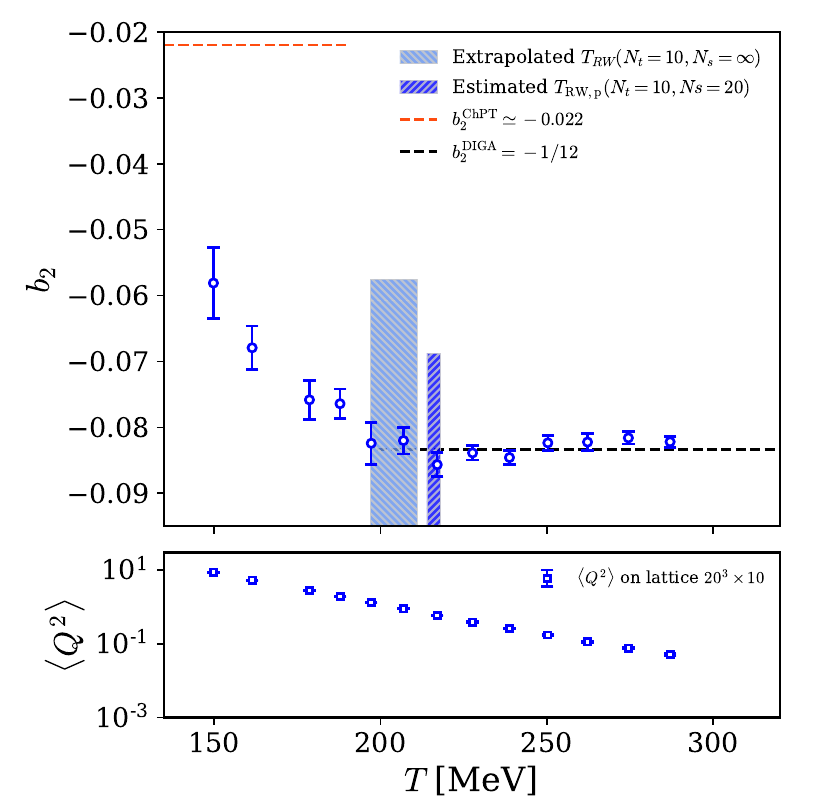}
  \caption{The fourth-order cumulant $b_2$ and $\braket{Q^2}$ as functions of temperature, computed on the $20^3 \times 10$.
    The vertical bands indicates $T_{RW}$ extrapolated to the thermodynamical limit, together
      with an estimate on the same lattice adopted for this measurement,
      while the horizontal dashed lines show the predicted values
    of the cumulant at low and high temperatures, respectively by ChiPT and DIGA. $\braket{Q^2}$ is here shown on a logarithmic scale.}
  \label{fig:b2_nt10}
\end{figure}

Figs.~\ref{fig:b2_nt8} and \ref{fig:b2_nt10} show the result obtained respectively on the  $16^3 \times 8$
and on the $20^3 \times 10$. The reported data correspond to accumulated statistics of $O(10^6)$ trajectories of unit length, with measurement taken every trajectory. A larger
statistics was used in correspondence of the lower temperatures, where the larger fluctuations make the
observable noisier.
Part of the results are quite clear: $b_2$ becomes compatible with the DIGA prediction as soon
as $T \geq T_{RW}$; this is true both for $N_t = 8$ and $N_t = 10$, so the result seems stable
towards the continuum limit. This is a behaviour similar to that observed in the pure gauge theory~\cite{Bonati:2013tt},
while a milder approach to the DIGA value has been reported along the standard thermal
line~\cite{Bonati:2015vqz,Kotov:2025ilm}, although with
statistical uncertainties larger than in the present investigation. The different behaviour could be related
to the fact that in this case we are dealing with a sharp transition, moreover associated with the spontaneous
breaking of a residual center symmetry, as for the pure gauge theory.

It is important
to assess the impact that finite size effects may have on these findings: indeed, it is
well known that on a small enough volume the topological charge distribution will eventually
be restricted to $Q=0$ and rare $|Q| = 1$ events only, which mimic a DIGA behavior in a fake way.
The way to check if this is the case is to measure $\langle Q^2 \rangle$ at the same time,
since the fake DIGA behaviour emerges when $\braket{Q^2} \ll 1$. This quantity is also reported in
Figs.~\ref{fig:b2_nt8} and \ref{fig:b2_nt10}, and shows that this kind of finite size effect
is indeed important for $T \gtrsim 250$~MeV, while determinations right above $T_{RW}$, which
are the most relevant to our conclusions, are safe.

Regarding the region below $T_{RW}$, we observe a decrease of $|b_2|$, which slowly approaches but never
reaches the ChiPT prediction in the explored range, which extends down to 150~MeV; morever, results obtained for
$N_t = 10$ differ from those on $N_t = 8$. In this case, finite size effects might be non-trivial.
Indeed, below $T_{RW}$ chiral symmetry is spontaneously broken and the inverse pion
mass sets the largest physical scale, so that a spatial size $N_s a \simeq 2$~fm, like we have
around $T_{RW}$ on our lattices with aspect ratio 2, could be too small, leading to sizeable deviations
from the ChiPT prediction. Moreover, because of the taste-breaking effects of the staggered discretization,
the pion spectrum depends on the lattice spacing, and this could explain the differences observed
between $N_t = 8$ and $N_t = 10$ also in terms of different finite size effects in the two cases.

\section{Conclusions}

\label{sec_conclusions}

In this study, we have investigated some properties of $N_f = 2+1+1$ QCD
and physical quark masses along the RW line, which is a modification of the standard
thermal line along which an exact residual center symmetry exists and gets spontaneously
broken at a critical temperature $T_{RW}$.
We have adopted a stout-smeared staggered discretization of the theory and,
as a first part of our investigation, we have checked that results obtained
for $T_{RW}$ and for the universality class of the transition are consistent
with those obtained for $N_f = 2+1$ QCD in Ref.~\cite{Bonati:2016pwz}.

We have then studied the topological properties of the theory along the RW line,
in order to highlight possible differences or analogies with respect to has been
observed in previous investigations along the standard thermal line. To this purpose,
we have measured the topological susceptibility and the fourth-order cumulant $b_2$ of the
topological charge distribution, which fix the first two coefficient of the Taylor expansion
of the free energy density $f(T,\theta)$ around $\theta = 0$.

As already observed in previous investigations of the thermal theory, both in pure gauge 
and full QCD, the topological susceptibility appears to be practically constant for $T \lesssim T_{RW}$,
then rapidly decaying for higher temperatures. Such decay is compatible with a power law dependence on $T$,
however with a power coefficient smaller that predicted by semiclassical approximation: this could be due to the slow
approach to the continuum limit, which indeed affected similarly also previous investigations along the standard thermal
line.

Results on $b_2$ instead show that it becomes compatible with the DIGA prediction as soon
as $T \geq T_{RW}$. This sharp transition is similar to what observed in the pure gauge theory~\cite{Bonati:2013tt},
while a milder approach was reported, although with a lower statistics, along the standard thermal
line for full QCD~\cite{Bonati:2015vqz,Kotov:2025ilm}.
The different behaviour is likely related 
to the fact that in this case we are dealing with a sharp transition; however, this suggests for the future
to increase the precision of the determination along the standard thermal line, in order to permit a
better comparison.

Below $T_{RW}$, $b_2$ shows a very slow approach to the ChiPT prediction, however in this case
our results could be affected by the fact that our lattice sizes are only slightly larger than
the pion Compton wavelength, leading to sizeable deviations from ChiPT predictions. The moderated
lattice sizes used for the determination of $b_2$ have been chosen to cope with the large
increase of the noise/signal ratio for this observable  as the volume increase. A possible way out,
in the future, could be to extend the techniques of analytic continuation from imaginary-$\theta$
to the full QCD case.

\acknowledgements

This work has been partially supported
by the project “Non-perturbative aspects of fundamental interactions, in the Standard Model and beyond” funded by MUR,
Progetti di Ricerca di Rilevante Interesse Nazionale (PRIN), Bando 2022, grant 2022TJFCYB (CUP I53D23001440006).
Numerical simulations have been performed on Leonardo at CINECA, based on the agreement between INFN and CINECA under projects INF25\_npqcd and INF26\_npqcd. KZ acknowledges support by the
project “Non-perturbative aspects of fundamental interactions, in the Standard Model and beyond” funded by MUR,
Progetti di Ricerca di Rilevante Interesse Nazionale (PRIN), Bando 2022, Grant 2022TJFCYB (CUP I53D23001440006).

\bibliography{biblio}

\end{document}